\documentclass[12pt]{iopart}
\pdfoutput=1
\usepackage{amssymb}
\usepackage{amstext}
\usepackage{enumerate}
\usepackage{relsize}
\renewcommand\theequation{\thesection.\arabic{equation}}
\usepackage{graphicx}
\usepackage[T1]{fontenc}
\usepackage[applemac]{inputenc}

\newcommand{\Z}{\mathbb Z}
\newcommand{\N}{\mathbb N}
\newcommand{\A}{\mathcal A}
\newcommand{\G}{\mathcal G}

\newtheorem{theo}{Theorem}[section]

\newcommand{\mc}[1]{{\mathcal #1}}

\newcommand{\mas}[1]{{\mathsmaller #1}}

\begin{document}

\title{Relative entropy via 
non-sequential recursive pair substitution}

\author{D. Benedetto$^1$, E. Caglioti$^1$, G. Cristadoro$^2$, 
M. Degli Esposti$^2$}
\address{$^1$  Dipartimento di Matematica, Universit\`a di Roma 
``La Sapienza'',
P.le A. Moro 5, 00185 Roma, Italy. e-mail:
benedetto@mat.uniroma1.it ; caglioti@mat.uniroma1.it}
\address{$^2$ Dipartimento di Matematica, 
Universit\`a di Bologna, P.za di Porta San Donato 5, 40126
Bologna, Italy. e-mail: desposti@dm.unibo.it ; cristadoro@dm.unibo.it}

\begin{abstract}
  The entropy of an ergodic source is the limit of properly rescaled
  1-block entropies of sources obtained applying successive
  non-sequential recursive pairs
  substitutions 
  \cite{G},\cite{bcg}. In this paper we prove that the cross entropy
  and the Kullback-Leibler divergence 
  can be obtained in a similar way.
\end{abstract}



\section{Introduction}\label{sec:intro}
 
Kullback-Leibler (KL) divergence (relative entropy) can be considered
as a measure of the difference/dissimilarity between
sources. Estimating KL divergence from finite realizations of a
stochastic process with unknown memory is a long-standing problem,
with interesting mathematical aspects and useful applications to
automatic categorization of symbolic sequences. Namely, an empirical
estimation of the divergence can be used to classify sequences (for
approaches to this problem using other methods, in particular true
metric distances, see \cite{lclmv}, \cite{os}; see also \cite{bbcd}).

In \cite{mz} Ziv and Merhav showed how to estimate the KL divergence
between two sources,
using the parsing scheme of LZ77 algorithm \cite{lz77} on
two finite length realizations.
They proved the consistence of the method by showing that the estimate of the divergence for two markovian sources  converges to their relative entropy 
when the length of the sequences diverges. Furthermore
they proposed this estimator as a tool for an ``universal 
classification'' of sequences.

A  procedure  based on the implementations
of LZ77 algorithm (gzip, winzip) is proposed in \cite{bcl}. The estimate obtained
of the relative entropy is then used to construct
phylogenetic trees for languages and is proposed as a tool to solve 
authorship attribution problems.
Moreover, the relation between the 
relative entropy and the estimate given by this procedure is
analyzed
in \cite{pbclv}.

Two different algorithms  are proposed and analyzed in \cite{ckv}, see also \cite{ckv2}. The first one is
based on the Burrows-Wheeler block sorting transform \cite{BWT}, while the other
uses the Context Tree Weighting method.  The authors proved the consistence of these approximation methods
and show that these methods outperform  the others  in  experiments.

\vskip3pt
In \cite{bcg} it is shown how to construct an entropy estimator
for stationary ergodic stochastic sources using non-sequential
recursive pairs substitutions method, introduced in \cite{G} (see also \cite{JLAM} and references therein for similar approaches).

In this paper we want to discuss the use of similar techniques to construct an  
estimator of relative (and cross) entropy 
between a pair of stochastic sources. In particular we investigate  how the 
asymptotic properties of concurrent pair substitutions might be used to construct
an optimal (in the sense of convergence) relative entropy estimator. 
A second relevant question arises about the computational efficiency
of the derived indicator. While here we address the first, mostly mathematical, 
question,  we leave the computational and
applicative aspects for forthcoming research.

\vskip3pt

The paper is structured as follows: in section \ref{sec:notations}
we state the notations,  in section \ref{sec:nsrps} we describe
the details of the non-sequential recursive pair substitutions (NSRPS) method, in section 
\ref{sec:scaling} we prove that NSRPS preserve the cross and the 
relative entropy, in section \ref{sec:convergence} we prove the main 
result: 
we can obtain an estimate of the relative entropy  by calculating
the 1-block relative entropy of the sequences we obtain 
using the NSRPS method.

\section{Definitions and notations}\label{sec:notations}
We introduce here the main definitions and notations, often following
the formalism used in \cite{bcg}.  Given a finite alphabet $\A$, we
denote with $\A^*=\cup_{k \geq 1}\A^k$ the set of finite words.  Given
a word $\omega\in \A^n$, we denote by $\vert\omega\vert=n$ its length
and if $1\leq i<j\leq n$ and
$\omega=(\omega_1,\omega_2,\ldots,\omega_n)$, we use $\omega_i^j$ to
indicate the subword $\omega_i^j=(\omega_i, \ldots,\omega_j)$. We use
similar notations for one-sided infinite (elements of $\A^{\mas \N}$)
or double infinite words (elements of $\A^{\mas \Z}$). Often sequences
will be seen as finite or infinite realizations of discrete-time
stochastic stationary, ergodic processes of a random variable $X$ with
values in $\A$.

The $n$-th order joint distributions $\mu_n$ identify the process and
its elements follow the consistency conditions:
$$
\mu_n(\omega_1^n)=
\sum_{\omega_0\in\A}\mu_{n+1}
(\omega_0,\omega_1,\ldots,\omega_n)=
\sum_{\omega_{n+1}\in\A}\mu_{n+1}(\omega_1,\ldots,\omega_n,\omega_{n+1}).
$$
When no confusion will arise, the subscript $n$ will be omitted, and
we will just use $\mu(\omega_1^n)$ to denote both the measure of the
cylinder and the probability of the finite word.

Equivalently, a distribution of a process can also be defined by
specifying the initial one-character distribution $\mu_1$ and the
successive conditional distributions:
$$
\mu(\omega_n\vert\omega_1^{n-1}) 
=\frac{\mu_n(\omega_1^n)}{\mu_{n-1}(\omega_1^{n-1})}.
$$
Given an ergodic, stationary stochastic source 
we define as usual:

\begin{description}
\item[\it n-block entropy]
$$
 H_n(\mu):= -\sum_{\vert\omega\vert=n}\mu(\omega)\log\mu(\omega).
$$
\item[\it n-conditional entropy]
$$h_n(\mu):= H_{n+1}(\mu)-H_n(\mu)=\sum_{\omega_1^n\in\A^n, a\in\A}\mu(\omega_1^na)\log\mu(a\vert w_1^n) := E_{\mu_{n+1}}\left(\log\mu(a\vert\omega_1^n)\right),
$$
where $\omega_1^na$ denotes the concatenated word $(\omega_1,\omega_2,\ldots,\omega_n,a)$ and $E_\mu\left(\bullet\right)$ is just the process average.
\item[\it Entropy of $\mu$]
$$
h(\mu):=\lim_{n\to\infty} \frac{H_n(\mu)}{n}=\lim_{n\to\infty} h_n(\mu)=E_\mu\left(\log\mu(a\vert\omega_1^\infty)\right)
$$
\end{description}

The following properties and results are very well known
\cite{shields}, but at the same time quite important for the proofs
and the techniques developed here (and also in \cite{bcg}):
\begin{itemize}
\item
$
h(\mu)\leq\ldots\leq 
h_k(\mu)\leq h_{k-1}(\mu)\leq\ldots\leq h_1(\mu)\leq H_1(\mu).$
\item 
A process $\mu$ is $k$-Markov if and only if $h(\mu)=h_k(\mu)$.
\item {\it Entropy Theorem}: 
for almost all realizations of the process, we have
$$
h(\mu)=\lim_{n\to\infty}-\frac{1}{n}\log\mu(x_1^n),
\qquad \mu-\text{a.s.}
$$
\end{itemize}

In this paper we focus on properties involving pairs of stochastic
sources on the same alphabet with distributions $\mu$ and $\nu$,
namely {\it cross entropy} and the related {\it relative entropy} (or
{\it Kullback Leibler divergence}):
 
\noindent
{\it n-conditional cross entropy}
\begin{equation}
\label{hk}
h_n(\mu || \nu) = 
-\sum_{{\omega} \in A^n,\,a\in A} \mu(
{\omega} a) \log \nu(a\vert\omega),
\end{equation}
{\it cross entropy}
\begin{equation}
\label{hktoh}
h(\mu || \nu) = 
\lim_{n\to +\infty}  h_n(\mu ||\nu),  
\end{equation}
{\it relative entropy (Kullback-Leibler divergence)}
\begin{eqnarray}
d(\mu\vert\vert\nu)&=
\lim_{n\to\infty} E_\mu\left(\log\frac{\mu(\omega_n\vert 
\omega_1^{n-1})}{\nu(\omega_n\vert \omega_1^{n-1})}\right)\nonumber
\\
&=\lim_{n\to\infty}\sum_{\omega_1^n\A^n} \mu(\omega_1^n)\log\frac{\mu(\omega_n\vert \omega_1^{n-1})}{\nu(\omega_n\vert \omega_1^{n-1})}.
\end{eqnarray}
Note that 
$$
h(\mu\vert\vert\nu) = h(\mu) + d(\mu\vert\vert\nu)
$$
Moreover we stress that, if
$\nu$ is k-Markov  then, for any $\mu$
\begin{equation}
\label{h-k-markov}
h(\mu \| \nu) = h_k(\mu \| \nu)
\end{equation}
Namely $h_l(\mu\|\nu)=h_k(\mu\|\nu)$ for any $l\ge k$:
$$
\begin{array}{ll}
h_l(\mu || \nu) &= - \sum_{\omega \in A^l,\,a\in A} \mu(
\omega a) \log \nu(a\vert\omega)
= - \sum_{\omega \in A^{l-k},\,b\in A^k,\,a\in A} \mu(
\omega ba) \log \nu(a\vert\omega b) \\[5pt]
&= - \sum_{\omega \in A^{l-k},\,b\in A^k,\,a\in A} \mu(
\omega ba) \log \nu(a\vert b) \\
&= 
 - \sum_{b\in A^k,\,a\in A} \mu(ba) \log \nu(a\vert b)= h_k(\mu\|\nu)
\end{array}
$$
Note that $h_1(\mu\|\nu)$ depends only 
on the two-symbol distribution of $\mu$. 

\vskip5pt

Entropy and cross entropy can be related to the asymptotic behavior of
properly defined {\it returning times} and {\it waiting times},
respectively.  More precisely, given an ergodic, stationary process
$\mu$, a sample sequence $w=w_1,w_2,\ldots$ and $n\geq 1$, we define
the returning time of the first $n$ characters as:

\begin{description}
\item[\it returning time]
\begin{equation}
R(w_1^n)\,=\,\min\{k> 1:\, w_{k}^{k+n-1}= w_1^n\}
\end{equation}
\end{description}
Similarly, given  two realizations $w = w_1,\ldots,w_n,\ldots$ 
and $z=z_1,\ldots,z_n,\ldots $ of $\mu$ and $\nu$ respectively, 
we define the 
\begin{description}
\item[\it waiting time]
\begin{equation}
W(w_1^n,z)=\min \{k>1:\, z_k^{k+n-1}=w_1^n\}
\end{equation}
\end{description}
Obviously  $W(w_1^n,w)=R(w_1^n)$.

We now have the following two important results:
\begin{theo}[Entropy and returning time \cite{ow}]
\label{returning}
If $\mu$ is a stationary, ergodic process, then 
$$
\lim_{n\to\infty}\frac{1}{n}\log R(w_1^n) =h(\mu)\quad \quad \mbox{$\mu-$a.s.}
$$
\end{theo}
\begin{theo}[Relative entropy and waiting time \cite{konto}]
\label{waiting}
If $\mu$ is stationary and ergodic, $\nu$ is k-Markov and 
the marginals $\mu_n$ of $\mu$ are dominated by the 
corresponding marginals $\nu_n$ of $\nu$, i.e. 
$\mu_n<<\nu_n$, then
$$
\lim_{n\to\infty}\frac{1}{n}\log W(w_1^n,z)= h(\mu) + d(\mu\vert\vert\nu)=h(\mu\vert\vert\nu),\qquad \mbox{$(\mu\times\nu)-$a.s.}
$$
\end{theo}

\section{Non sequential recursive pair substitutions}
\label{sec:nsrps}
We now introduce a family of transformations on sequences and the
corresponding operators on distributions: given $a,b\in\A$ (including
$a=b$), $\alpha\notin\A$ and $\A'=\A\cup\{\alpha\}$, a {\it pair
  substitution} is a map $G_{ab}^\alpha:\A^*\to {\A}^{'*}$ which
substitutes sequentially, from left to right, the occurrences of $ab$
with $\alpha$. For example
$$
G_{01}^2\left(0010001011100100\right)=020022110200.
$$
or:
$$
G_{00}^2(0001000011)=2012211.
$$ 
$G=G_{ab}^\alpha$ is always an injective but not surjective map
that can be immediately extended also to infinite sequences
$w\in\A^{\mas \N}$.

The action of $G$ shorten the original sequence: we denote by $Z$ the
inverse of the contraction rate:
$$
\frac{1}{Z_{ab}(\omega_1^n)}:=\frac{\vert G_{ab}^{\alpha}(\omega_1^n)\vert}{\vert\omega_1^n\vert}=1-\frac{\sharp\{ab\subseteq\omega_1^n\}}{n}
$$
For $\mu$-{\it typical} sequences we can pass to the limit and define:
\begin{equation*}
\frac{1}{Z^\mu}:=\lim_{n\to\infty}
\frac{\vert G(\omega_1^n)\vert}{\vert\omega_1^n\vert}=\left\{ 
\begin{array}{ll}
 1-\mu(ab) &  \text{if } a\neq b \\
 1-\mu(aa)+\mu(aaa)-\mu(aaaa)+
\cdots& \text{if } a=b
\end{array}
 \right.
 \end{equation*}

An important remark is that if we start from a source where
admissible words are described by constraints on consecutive symbols,
this property will remain true even after an arbitrary pair
substitution. In other words (see Theorem 2.1 in \cite{bcg}): a pair
substitution maps pair constraints in pair constraints.

A pair substitution
$G_{ab}^{\alpha}$
naturally induces a map on the set of ergodic stationary measures on
$A^{\mas\Z}$ by mapping typical sequences w.r.t. the original
measure $\mu$ in typical sequences w.r.t. the transformed measure
$\G\mu$: given $z_1^k\in{\A'}^{*}$ then (Theorem 2.2 in \cite{bcg})
$$\G
\mu(z_1^k):=\lim_{n\to\infty}\frac{\sharp\{z_1^k\subseteq
  G(\omega_1^n)\}}{\vert G(\omega_1^n)\vert}
$$
exists and is constant $\mu$ almost everywhere in
$\omega\in\A^{\mas \N}$, moreover
$\{\G\mu(z)\}_{z\in{\A'}^{*}}$ are the marginals of an ergodic
measure on ${\A'}^{\mas\Z}$. 
 
Again in \cite{bcg} , the following results are proved showing how
entropies transform under the action of $\G=\G_{ab}^{\alpha}$, with
expanding factor $Z=Z_{ab}^\mu$:

\noindent
{\it Invariance of entropy}
$$
h(\G\mu)=Z\,h(\mu).
$$
{\it Decreasing of the 1-conditional entropy}
$$
h_1(\G\mu)\leq Z h_1(\mu).
$$
Moreover, $\G$ maps 1-Markov measures in 1-Markov measures. In fact:
$$
h(\G\mu)\leq h_1(\G\mu)\leq Z h_1(\mu)=Z h(\mu)=h(\G\mu)
$$
{\it Decreasing of the k-conditional entropy}
$$
h_k(\G\mu)\leq Z h_k(\mu).
$$
Moreover $\G$ maps $k$-Markov measures in $k$-Markov measures.

\vskip3pt

While later on we will give another proof of the first fact, we remark
that this property, together with the decrease of the 1-conditional
entropy, reflect, roughly speaking, the fact that the amount of
information of $G(\omega)$ , which is equal to that of $\omega$, is
more concentrated on the pairs of consecutive symbols.

As we are interested in sequences of recursive pair substitutions, we
assume to start with an initial alphabet $\A$ and 
define an increasing alphabet sequence
$\A_1$, $A_2$, \dots $\A_{N}$, \dots.
Given $N$ and chosen $a_{\mas N},\,b_{\mas N}\in \A_{N-1}$
(not necessarily different):

\begin{itemize}
\item[] 
we indicate with $\alpha_{\mas N}\notin A_{N-1}$ a new symbol
and define the new alphabet as 
$\A_N = \A_{N-1} \cup \{ \alpha_{\mas N} \}$;
\item[]
we denote with $G_N$ the substitution map 
$G_N= G_{a_{\mas N}b_{\mas N}}^{\alpha_{\mas N}}:
\A_{N-1}^*\to\A_N^*$
which substitutes whit $\alpha_{\mas N}$
the occurrences of the pair  
$a_{\mas N}b_{\mas N}$
in the strings on the alphabet $\A_{N-1}$;
\item[]
we denote with $\G_N$  the corresponding map from the measures
on $A_{N-1}^\Z$ to the measures on $A_N^\Z$;
\item[]
we define by $Z_N$ the corresponding normalization
factor $Z_N = Z_{a_{\mas N}b_{\mas N}}^{\alpha_{\mas N}}$.
\end{itemize}

\noindent
We use the over-line to denote iterated quantities:
\begin{eqnarray*}
\bar{G}_{\mas N}&:=
&G_{\mas N}\circ G_{\mas N-1}\circ\cdots\circ G_1,
\qquad\quad  \bar{\G}_{\mas N}:=\G_{\mas N}\circ \G_{\mas N-1}\circ\cdots\circ \G_1
\end{eqnarray*}
and also
$$
\bar{Z}_{\mas N}=Z_{\mas N}Z_{\mas N-1}\cdots Z_1.
$$

The asymptotic properties of $\bar{Z}_{\mas N}$ clearly depend on the
pairs chosen in the substitutions. In particular, if at any step $N$
the chosen pair $a_{\mas N}b_{\mas N}$ is the pair of maximum of
frequency 
of $\A_{\mas N-1}$ then (Theorem 4.1 in \cite{bcg}):
$$
\lim_{N\to\infty}\bar{Z}_{\mas N}=+\infty
$$

Regarding the asymptotic properties of the entropy we have the
following theorem that rigorously show that $\mu_{\mas N}:=\bar{G}_N\mu$ 
becomes asymptotically 1-Markov:
\begin{theo}[Entropy via NSRPS \cite{bcg}] If
$$
\lim_{N\to\infty}\bar{Z}_{\mas N}=+\infty
$$
then
$$
h(\mu)=\lim_{N\to\infty} \frac{1}{\bar{Z}_{\mas N}} h_1(\mu_{\mas N})
$$
\end{theo}
The main results of this paper is the 
generalization of this theorem to the cross and relative entropy.

Before entering in the details of our construction let us sketch here the main steps.

In particular let us consider the cross entropy (the same argument will apply to the relative entropy) of the measure $\mu$ with respect to the measure $\nu$:  i.e. $h(\mu\vert\vert \nu)$.
 
As we will show, but  for the normalization factor ${\overline Z}_{\mas N}^{\mu}$, this is equal to the cross entropy of  the measure $G_{N}\mu$ w.r.t the measure $G_{N}\nu$:
$$h(\mu||\nu) =
\frac {h(\G_{\mas N}\mu||\G_{\mas N}\nu)}
{{\overline Z}_{\mas N}^{\mu}}$$

Moreover, as we have seen above ,  if we choose the substitution in a suitable way (for instance if at any step we substitute the pair with maximum frequency) then ${\overline Z}_N^{\nu}\rightarrow\infty$ and the measure $G_{N}\nu$ becomes asymptotically 1-Markov as $N\rightarrow\infty$.

Interestingly, we do not  know if $Z_{\nu}$ also diverges (we will discuss this point in the sequel).

Nevertheless, noticing that  the cross entropy of  a 1-Markov source w.r.t a  generic ergodic source is equal to the 1-Markov cross entropy between the two sources, it is reasonable to expect that the cross entropy $h(\mu\vert\vert \nu)$
can be obtained as the following limit:

$$h(\mu||\nu) = \lim_{N\to +\infty} 
\frac {h_1(\G_{\mas N}\mu||\G_{\mas N}\nu)}
{{\overline Z}_{\mas N}^{\mu}}$$

This is exactly what we will prove in the two next sections.

\section{Scaling of (relative) entropy via waiting times}
\label{sec:scaling}
We first show how the relative entropy between two stochastic process
$\mu$ and $\nu$ scales after acting with the \emph{same} pair
substitution on both sources to have $\G \mu$ and $\G \nu$. More
precisely we make use of Theorem \ref{waiting} and have the following:

\begin{theo}[Invariance of relative entropy for pair substitution]
\label{main1} If $\mu$ is ergodic, 
$\nu$ is a Markov chain and $\mu_n<<\nu_n$, then
if $G$ is a pair substitution
$$d(\G\mu\vert\vert \G\nu)=Z^\mu d(\mu\vert\vert \nu)
$$
\end{theo}

{\it Proof.} 
To fix the notations, let us denote by $w$ and $z$ the infinite realizations
of the process of measure 
$\mu$ and $\nu$ respectively, and by $w_k^n$
and $z_k^n$ the corresponding finite substrings. Let us denote by
$a,b\in\A$ the characters involved in the pair substitution
$G=G_{ab}^\alpha$. 
Moreover let us denote the waiting time with the shorter notation:
$$t_{n}:=W(w_1^n , z).$$
We now explore how the waiting time rescale with respect to the
transformation $G$: we consider the first time we see the sequence
$G(w_1^n)$ inside the sequence $G(z)$.  To start with, we assume
that $w_1 \neq b$ as we can always consider 
Th. \ref{waiting} for realizations with a fixed prefix
of positive probability. Moreover
we choose a subsequence $\{n_i \}$ such that $n_i$ is the smallest
$n>n_{i-1}$ such that $w_{n_i}\neq a$. Of course $n_i \to \infty $ as
$i\to \infty$. 
In this case, it is easy to observe that
\begin{equation*}
W(G(w_1^{n_i}), G(z)) = |G(w_1^{t_{n_i}})|
\end{equation*}
Then, using Theorem \ref{waiting}
\begin{eqnarray}
h(\G \mu\|\G \nu) 
&=& \lim_{i\to +\infty} \frac{1}{|G(w_1^{n_i})|}\log\left[ W(G(w_1^{n_i}), G(z)) \right] =\nonumber\\
&=& \lim_{i\to +\infty} \frac{n_i}{|G(w_1^{n_i})|}\frac{\log|G(w_1^{t_{n_i}})|}{n_i}= \nonumber\\
&=& \lim_{i\to +\infty} \frac{n_i}{|G(w_1^{n_i})|}\left[\frac{1}{n_i}\log( t_{n_i} ) +\frac{1}{n_i}\log\left(\frac{|G(w_1^{t_{n_i}})|}{t_{n_i}}\right)\right]=\nonumber\\
&=& Z^{\mu} h(\mu\|\nu) \label{KL1}
\end{eqnarray}
where in the last step we used the fact that $t_{n_i}\to \infty$ as
$i\to \infty$, the definition of $Z^\mu$ and Theorem \ref{waiting} for
$\mu$ and $\nu$.  Note that for $\mu = \nu$, equation (\ref{KL1})
reproduces the content of Theorem 3.1 of \cite{bcg}:
$$
h(\G\mu)=Z^{\mu}h(\mu),
$$
that thus implies
$$
d(\G\mu||\G\nu)=Z^{\mu} d(\mu||\nu).
$$
Note that the limit in Th. \ref{waiting} 
is almost surely unique and then 
the initial restrictive assumption $w_1\neq b$ and the use of
the subsequence $n_i$ have no consequences
on the thesis; this  concludes
the proof.

$\Box$

\vskip3pt
Before discussing the convergence of relative entropy under successive
substitutions we go thorough a simple explicit example of the Theorem
\ref{main1}, in order to show the difficulties we deal with, 
when we try to use the explicit expressions of the transformed measures
we find in \cite{bcg}.

\vskip3pt
\noindent
{\it Example.} We treat here the most simple case: $\mu$ and $\nu$ are
  Bernoulli binary processes with parameters $\mu(0),\mu(1)$ and
  $\nu(0),\nu(1)$ respectively.  We consider the substitution
  $G=G_{01}^2$ given by $01\to2$.  It is long but easy to verify that  $\G\mu$ is a stationary, ergodic, 1-Markov with equilibrium
  state
$$
\G\mu(0)=Z\mu(00),\quad \G\mu (1)=Z(\mu(1)-\mu(01)),\G\mu(2)=Z\mu(01)\phantom{a},
$$
where $Z=Z^\mu(01)=(1-\mu(01))^{-1}$.

For example, given a $\G\mu$-generic sequence $y_1,\ldots,y_m$, corresponding to a $\mu$-generic sequence $x_1,\ldots,x_n$   ($y=Gx$):
\begin{eqnarray*}
\G\mu(0)&=&\lim_{m\to\infty}\frac{1}{m}\sharp\{0\in y_1^m\}\\
&=&\lim_{n\to\infty}\frac{n}{m}\cdot\frac{\sharp\{0\in x_1^n\}-\sharp\{01\in x_1^n\}}{n}\\
&=&(\mu(0)-\mu(01))\cdot\lim_{n\to\infty}\frac{n}{n-\sharp\{01\in x_1^n\}}= Z(\mu(0)-\mu(01))= Z \mu(00)
\end{eqnarray*}
Clearly:
$$
\G\mu(0)+\G\mu(1)+\G\mu(2)=1$$

Using the same argument as before, it is now possible to write down
the probability distribution of pair of characters for $\G\mu$. Again
the following holds for a generic process:
{
\footnotesize
$$
\begin{array}{lll}
\frac{\G\mu(00)}Z=\mu(00)- \mu(001) = \mu(000)& 
\frac{\G\mu(01)}Z= 0 &
\frac{\G\mu(02)}Z= \mu(001)\\[4pt]
\frac{\G\mu(10)}Z= \mu(10)-\mu(010) -\mu(101)+\mu(0101) & 
\frac{\G\mu(11)}Z= \mu(11)-\mu(011)& 
\frac{\G\mu(12)}Z= \mu(101)-\mu(0101)\\[4pt]
\frac{\G\mu(20)}Z= \mu(010)-\mu(0101)& 
\frac{\G\mu(21)}Z= \mu(011) &
\frac{\G\mu(22)}Z= \mu(0101)
\end{array}
$$
}

It is easy to see that $\sum_{x,y=0,1,2} \G\mu(xy)=1$. Now we can
write the transition matrix $P$ for the process $\G\mu$ as
$P(y\vert x)={\G\mu(xy)}/{\G\mu(x)}$:
$$
P =   \left(\begin{array}{ccc}
    P(0\vert 0)& P(1\vert 0) & P(2\vert 0) \\ 
     P(0\vert 1)& P(1\vert 1) & P(2\vert 1) \\ 
    P(0\vert 2)& P(1\vert 2) & P(2\vert 2) \\ 
  \end{array}\right)\\
$$
For Bernoulli processes:
$$
P=  \left(\begin{array}{ccc}
\mu(0)& 0 &  \mu(1)\\ 
  \mu(00)&  \mu(1) &   \mu(01)\\ 
\mu(00)& \mu(1)&  \mu(01)\\ 
  \end{array}\right).
$$

We now denote with $Q$ the transition matrix for $Q\nu$. For the two
1-Markov processes, we have
$$
d(\G\mu\vert\vert \G\nu)=
\sum_{x=0,1,2}\mu(x)
\sum_{y=0,1,2} P(y\vert x)\log\frac{P(y\vert x)}{Q(y\vert x)}.
$$
Via straightforward calculations, using the product structure of the
measure $\mu$:
\begin{eqnarray*}
d(G\mu\vert\vert G\nu)= 
Z\mu(00)\left[\mu(0)\log\frac{\mu(0)}{\nu(0)}+\mu(1)\log\frac{\mu(1)}{\nu(1)}\right]\\
+ Z\mu(11)\left[\mu(00)\log\frac{\mu(00)}{\nu(00)}+\mu(1)\log\frac{\mu(1)}{\nu(1)}+\mu(01)\log\frac{\mu(01)}{\nu(01)}\right]\\
+ Z\mu(01)\left[\mu(00)\log\frac{\mu(00)}{\nu(00)}+\mu(1)\log\frac{\mu(1)}{\nu(1)}+\mu(01)\log\frac{\mu(01)}{\nu(01)}\right]\\
= Z\mu(00) d(\mu\vert\vert\nu)+ Z\mu(1)\left[\mu(00)\log\frac{\mu(00)}{\nu(00)}+\mu(1)\log\frac{\mu(1)}{\nu(1)}+\mu(01)\log\frac{\mu(01)}{\nu(01)}\right]\\
=  Z\mu(00) d(\mu\vert\vert\nu)+Z\mu(1)\left[\mu(0)d(\mu\vert\vert\nu)+ d(\mu\vert\vert\nu)\right]\\
=Z d(\mu\vert\vert\nu) (\mu(00)+\mu(10)+\mu(1))\\
= Z d(\mu\vert\vert\nu)
\end{eqnarray*}

\section{The convergence}
\label{sec:convergence}

We now prove that the renormalized 1-Markov cross entropy between $\mu_n$ and $\nu_n$ converges to the cross-entropy between $G_{n}\mu$ and $G_{n}\nu$ as the number of pair substitution $n$ goes to $\infty.$. 

More precisely:
\begin{theo}[KL divergence via NSRPS]
\label{main2} If ${\overline Z}_{\mas N}^{\nu}
\to +\infty$ as $N\to +\infty$,
$$h(\mu||\nu) = \lim_{N\to +\infty} 
\frac {h_1(\G_{\mas N}\mu||\G_{\mas N}\nu)}
{{\overline Z}_{\mas N}^{\mu}}$$
\end{theo}

{\it Proof.}
Let us define, as in \cite{bcg} the following operators on the ergodic measures:
$\mc P$ is the projection operator that maps
a measure to its 1-Markov approximation, whereas 
$\mc P_{\mas N}$ is the operator such that for any arbitrary $\nu$
$$
\mc P \bar{\mc G}_{\mas N}\nu = \bar{\mc G}_{\mas N} \mc P_{\mas N}\nu
$$
We notice (see \cite{bcg} for the details)  that 
the normalization constant for $\mc P_{\mas N}\nu$
is the same of that for $\nu$:
$$
Z_{\mas N}^{\nu} = Z_{\mas N}^{\mc P_{\mas N}\nu }.
$$

The measure $\mc P_{\mas N}\nu$ is not $1$-Markov, but we know that it
becomes 1-Markov after $N$ steps of substitutions, in fact it becomes
$\mc P \bar{G}_{\mas N}\nu$.  Moreover, as discussed in \cite{bcg}, it
is an approximation of $\nu$ if ${\overline Z}_{\mas N}^{\nu}$
diverges: for any $\omega$ of length $k$,
\begin{equation}
\label{convergenza}
|\mc P_{\mas N}\nu(\omega) - \nu(\omega)| \le 2\, \frac{k^2}
{{\overline Z}_{\mas N}^{\nu}}
\end{equation}
Now it is easy to establish the following chain of equalities:
$$h(\mu || \mc P_{\mas N} \nu ) = 
\frac 1{{\overline Z}_{\mas N}^{\mu}} h(\mc G_{\mas N} \mu ||
\mc G_{\mas N} \mc P_{\mas N} \nu ) = 
\frac 1{{\overline Z}_{\mas N}^{\mu}} 
h(\mc G_{\mas N} \mu || \mc P \mc G_{\mas N} \nu ) = 
\frac 1{{\overline Z}_{\mas N}^{\mu}} h_1(\mc G_{\mas N} \mu ||
\mc G_{\mas N} \nu )
$$
where we have used the conservation of  the cross entropy $h$  
and the fact that $H(\pi||\xi) = h_1(\pi||\xi)$ 
if $\xi$ are 1-Markov, as shown in eq. \ref{h-k-markov}.
To conclude the proof we have to show that 
$$h(\mu || \mc P_{\mas N} \nu )\to h(\mu||\nu)$$
This is an easy consequence of eq. \ref{convergenza}
the definition 
\ref{hk} and eq. \ref{hktoh}. $\Box$

\section{Conclusions and remarks}
\label{sec:conclusions}

It is important to remark that we are \underline{not} assuming the
divergence of $\bar{Z}^\mu_{\mas N}$ too, as not being necessary for the
convergence to the (rescaled) two-characters relative entropy.

\begin{figure*}
  \includegraphics[scale=1]{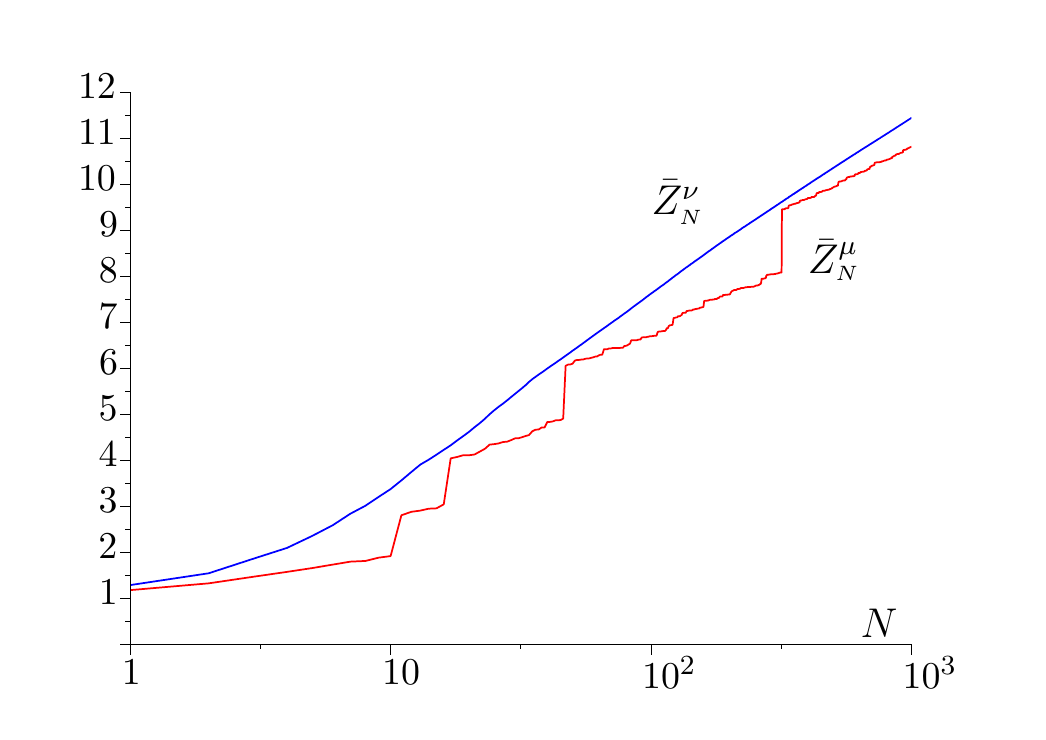} 
  \caption{$\bar Z_{\mas N}^{\nu}$ and $\bar Z_{\mas N}^{\mu}$ for
    realizations of $10^8$ characters of two Markov process on
    $\{0,1\}$ with memory of length 5.  The pairs are chosen with the
    standard strategy (most frequent pair substitution) for the
    sequences of starting measure $\nu$.  }
  \label{fig:Z}
\end{figure*}

Nevertheless, it would be interesting  to understand both the topological
and statistical constraints that prevent or permit the divergence of
the expanding factor $\bar{Z}^\mu_{\mas N}$.  Experimentally, it seems
that if we start with two measures with finite relative entropy
(i.e. with absolutely continuous marginals), then 
if we choose the standard strategy (most frequent pair substitution) for the sequence of pair substitutions
that yields the divergence of
$\bar{Z}_{\mas N}^\nu$,   we also simultaneously obtain the divergence of $\bar{Z}_{\mas N}^\mu$
(see for instance fig. \ref{fig:Z}).

On the other hand, it seems possible to consider particular
sources and particular strategies of pairs substitutions 
withdiverging $\bar{Z}_{\mas N}^\nu$, that
prevent the divergence of $\bar{Z}_{\mas N}^\mu$.
At this moment we do not have conclusive rigorous mathematical results on this subject.

Finally, let us note that  Th. \ref{main2} do not give 
directly an algorithm to estimate the relative entropy:
in any implementation we would have to specify the ``optimal'' number of pairs substitutions, 
with respect to the length of initial sequences and
also with respect to the dimension of the initial alphabet.
Namely, in the estimate we have to take into account at least two 
correction terms, which diverges with $N$: the 
entropy cost of writing the substitutions and 
the entropy cost of writing the frequencies of the pairs of characters in
the alphabet we obtain after the substitutions
(or equivalent quantities if we use, for instance, 
arithmetic codings modeling the two character frequencies). 

For what concerns possible implementations of the method it is important to notice that the NSRPS procedure can be implemented in linear time \cite{JLAM}.
Therefore it seems reasonable that reasonably fast algorithms to compute relative entropy via NSRPS can be designed.
Anyway, preliminary numerical experiments
show that for sources of finite memory this method
seems to have the same limitations 
of that based on parsing procedures, with respect
to the methods based on the analysis of context
introduced in 
\cite{ckv}.

\begin{figure*}
  \includegraphics[scale=1]{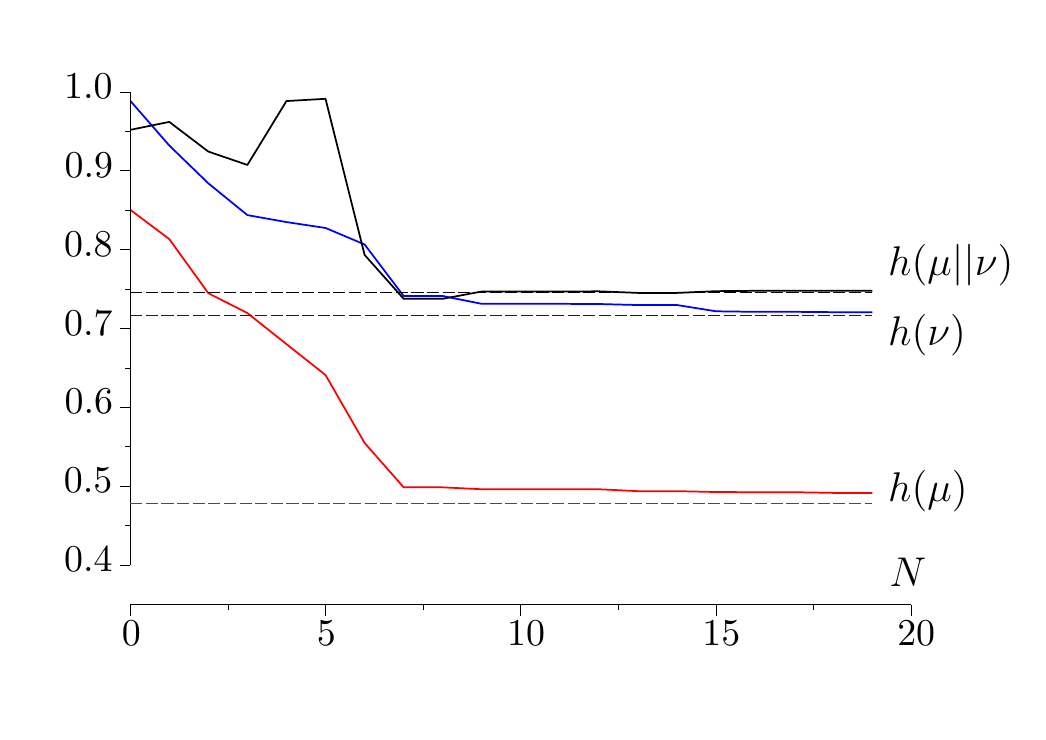} 
  \caption{Solid lines are the estimates of $h(\nu)$, $h(\mu)$ and of
    the cross entropy $h(\mu || \nu)$ obtained after $N$ pairs
    substitutions. The dashed lines are the corresponding analytic
    value.}
  \label{fig:H}
\end{figure*}

In fig. \ref{fig:H}
we show the convergence of the estimates of the entropies of the two
sources and of the cross entropy, given Th. \ref{main2},
for two Markov process of memory 5.
In this case, the numbers of substitutions $N=20$ is small
with respect to the length of the sequences $10^8$, then
the correction terms are negligible.

Let us finally note that the  cross entropy estimate  might  show large variations
for particular values of $N$. This could be interpreted by the fact that
for these values of $N$  pairs with particular relevance
for one source with respect to the other have been substituted.
This example suggest that the NSRPS method for the
estimation of the cross entropy should be useful in  sequences analysis, for example in order
to detect  strings with a peculiar statistical role.

\end{document}